\documentstyle[12pt]{article}\textheight
230mm\textwidth 165mm 
\newcommand{\bi}{\bibitem}
\newcommand{\be}{\begin{eqnarray}}
\newcommand{\ee}{\end{eqnarray}}
\newcommand{\nn}{\nonumber}
\catcode`\@=11
\def\lsim{\mathrel{\mathpalette\@versim<}}
\def\gsim{\mathrel{\mathpalette\@versim>}}
\def\@versim#1#2{\vcenter{\offinterlineskip
\ialign{$\m@th#1\hfil##\hfil$\crcr#2\crcr\sim\crcr } }}
\catcode`\@=12

\begin{document}

\pagestyle{empty}

\noindent
\hspace*{10.7cm} \vspace{-3mm}  HIP-1999-40/TH\\

\noindent
\hspace*{10.7cm} June 1999

\begin{center}
{\Large\bf Gauge Coupling Unification with Extra Dimensions 
and Correction due to Higher Dimensional Operators}
\end{center} 

\vspace{1cm}
\begin{center}
{\sc Katri Huitu}$\ ^{(1),\dag}$ and
{\sc Tatsuo Kobayashi}$\ ^{(1,2),\dag}$ 
\end{center}
\begin{center}
{\em $\ ^{(1)}$ 
Helsinki Institute of Physics, 
FIN-00014 University of Helsinki, Finland} \vspace{-2mm}\\
{\em $\ ^{(2)}$ 
Department of Physics,  
FIN-00014 University of Helsinki, Finland} \vspace{-2mm}\\
\end{center}

\vspace{1cm}
\begin{center}
{\sc\large Abstract}
\end{center}

\noindent
We study the gauge coupling unification with extra dimensions.
We take into account corrections due to the higher dimensional operators.
We show the prediction of $\alpha_3(M_Z)$ is sensitive to such corrections, 
even if $c<\Phi>/M=O(0.01)$.
We also discuss the $b-\tau$ Yukawa unification.


\vspace{4cm}
\footnoterule
\vspace{0.1cm}
\noindent
$^{\dag}$Partially supported  by the Academy of Finland \vspace{-3mm} 
(no. 44129).\\

\newpage
\pagestyle{plain}

Recently, theories with large extra dimensions have been studied 
intensively \cite{antoniadis4}-\cite{extra-ref2}.
If such extra dimensions correspond to a TeV scale that can be 
a solution of the naturalness problem.
One of interesting aspects in the theory with extra dimensions is 
the power-law behaviour of the running gauge coupling constants shown 
in Ref.~\cite{dienes1}, that is, 
the towers of Kaluza-Klein excitation modes lead to the power-law behaviour.
That provides with the possibility that the three gauge coupling 
constants of the standard model 
are unified at lower energy scale than $10^{16}$ GeV and 
the unified energy scale is just above 
the energy scale $\mu_0$ where the Kaluza-Klein excitation modes appear.

However, detailed analyses on the minimal matter content and canonical 
level of $U(1)_Y$
show that the predicted value of $\alpha_3(M_Z)$ increases as 
$\mu_0$ decreases from $10^{16}$ GeV to a TeV scale, and 
we obtain incorrect prediction for $\alpha_3(M_Z)$ \cite{GR,KKMZ}.
There are several works to obtain a realistic prediction of 
$\alpha_3(M_Z)$ leading to the experimental value \cite{p-data}, 
\begin{equation}
\alpha_3(M_Z) = 0.119 \pm 0.002,
\end{equation}
e.g. by considering the non-canonical 
level of $U(1)_Y$ different from 
$5/3$ or adding extra matter fields \cite{extra-uni,KKMZ} \footnote{See 
also Ref.~\cite{AK}.}.

In this paper we consider the case with the minimal matter content and 
the canonical level of $U(1)_Y$ equal to $5/3$, 
and we take into account the correction due to the higher dimensional 
operator, e.g. $c(\Phi/M)FF$, where $c$ is a coupling constant and 
$M$ is the cut-off energy scale.
We consider the $SU(5)$ grand unified theory (GUT) 
with the {\bf 24} Higgs field $\Phi$ as our framework.
Within this framework, we have the correction term 
to the gauge kinetic term \cite{g-nonuni} \footnote{In 
Ref.~\cite{HKKP}, corrections due to Higgs fields with larger 
representations are discussed within the framework of 
the four-dimensional supersymmetric $SU(5)$ GUTs.}, 
\begin{equation}
-{1 \over 4}c{<\Phi_{\alpha \beta}> \over M}F^{\alpha}_{\mu \nu}
F^{\beta \mu \nu}.
\end{equation}
Note that the vacuum expectation value 
\begin{equation}
<\Phi_{\alpha \beta}> \propto {\rm diag} (1/2\sqrt{15})(2,2,2,-3,-3),
\end{equation} 
which corresponds to the $SU(3) \times SU(2) \times U(1)_Y$ preserving 
direction, contributes to the gauge coupling constants non-universally.
Thus, the initial condition changes into  
\begin{eqnarray}
\alpha_i^{-1} &=& \alpha_X^{-1}(1+C_i), \\ 
(C_1,C_2,C_3) &=& {x \over 2\sqrt{15}}(-1,-3,2)
\label{in-uni}
\end{eqnarray}
where
$x=c<\Phi>/M$.
With this initial condition, let us study the gauge coupling unification 
with extra dimensions, that is, the prediction of $\alpha_3(M_Z)$.
We will also study the $b-\tau$ Yukawa unification.

First, we give the set-up of our model and its renormalization 
group (RG) equations  \cite{dienes1}.
Following ref. \cite{dienes1} we assume
that only the gauge boson and Higgs supermultiplets of 
the minimal supersymmetric standard model (MSSM) are in the bulk and 
have the towers of Kaluza-Klein states and that
the lepton and quark supermultiplets are sitting at a fixed point
of an orbifold on which the $\delta$ dimensional internal space is
compactified so that they have no  towers of Kaluza-Klein states.
It is easy to extend to the case that some quarks or leptons have 
the Kaluza-Klein towers.
Under these assumptions, 
the one-loop $\beta$-functions of the gauge couplings $g_i$ ($i=1,2,3$)
and the Yukawa couplings $g_{t,b,\tau}$ above $\mu_0$ 
become \cite{dienes1} \footnote{In Ref.\cite{KKMZ} $\beta$-functions 
of soft supersymmetry breaking parameters have also been obtained 
by use of the recently developed technique based on the spurion 
formalism \cite{jack3,kkz1}.}:
\be
(16 \pi^2)\beta_1&=& g_1^3~(6+\frac{6}{5}(Y_\delta/2)
(\frac{\Lambda}{\mu_0})^{\delta}), \\
(16 \pi^2)\beta_2&=&
g_2^3~(4-6(Y_\delta/2) (\frac{\Lambda}{\mu_0})^{\delta}), \\
(16 \pi^2)\beta_3&=&
g_3^3~(3-12(Y_\delta/2) (\frac{\Lambda}{\mu_0})^{\delta}),\\
(16 \pi^2)\beta_t&=&
g_t~[3 g_t^2-\frac{3}{10}g_1^2-\frac{3}{2} g_2^2\nn\\
& &+(Y_\delta/2)~ (\frac{\Lambda}{\mu_0})^{\delta}
(6g_t^2+2g_b^2
-\frac{17}{15}g_1^2-3g_2^2-\frac{32}{3}g_3^2)]~,\\
(16 \pi^2)\beta_b&=&
g_b~[3 g_b^2+g_\tau^2-\frac{3}{10}g_1^2-\frac{3}{2} g_2^2\nn\\
& &+ (Y_\delta/2)~ (\frac{\Lambda}{\mu_0})^{\delta}
(2g_t^2+6g_b^2
-\frac{1}{3}g_1^2-3g_2^2-\frac{32}{3}g_3^2)]~,\\
(16 \pi^2)\beta_\tau&=&
g_\tau~[3 g_b^2+g_\tau^2-\frac{3}{10}g_1^2-\frac{3}{2}g_2^2\nn\\
& &+ (Y_\delta/2) ~(\frac{\Lambda}{\mu_0})^{\delta}
(6g_\tau^2
-3g_1^2-3g_2^2)]~,
\ee
where 
we have neglected the Yukawa couplings of the
first and second generations, and 
$Y_\delta$ is defined as \cite{KKMZ}
\be
Y_\delta=\frac{\pi^{\delta/2}}{\Gamma(2+\delta/2)}~.
\label{Ydelta}
\ee
This coefficient corresponds to $X_\delta$ in Ref.~\cite{dienes1}, 
 \be
X_\delta=\frac{\pi^{\delta/2}}{\Gamma(1+\delta/2)}~,
\label{Xdelta}
\ee
but these are different each other by the factor $1+\delta/2$.
In contrast with Ref.~\cite{dienes1}, in Ref.~\cite{KKMZ} the 
matching condition between the four-dimensional effective theory and 
$D+\delta$ dimensional theory is required to obtain $Y_\delta$ 
such that the evolution equations of the couplings in the 
effective theory smoothly go over in the large compactification radius 
limit to those in the uncompactified, original, 
$D+\delta$ dimensional theory.
In particular, the continuous Wilson RG approach, which is applicable 
in any dimensions, is employed.

Below the energy scale $\mu_0$, we use the two-loop RG equations 
of the four dimensional MSSM.
For simplicity, we take $\delta =1$.
Under the initial condition (\ref{in-uni}),  we predict $\alpha_3(M_Z)$ 
using these RG equations with the experimental values \cite{p-data},
\begin{equation}
M_\tau=1.777 ~{\rm GeV}, \quad M_Z=91.188 ~{\rm GeV},
\end{equation}
\begin{equation}
\alpha^{-1}_{\rm EM}(M_Z) = 127.9 +{8 \over 9 \pi} 
\log {M_t \over M_Z}~,
\end{equation}
\begin{equation}
\sin ^2 \theta_W(M_Z) =0.2319 - 3.03\times 10^{-5} T-8.4 \times 
10^{-8}T^2~,
\end{equation}
where $T=M_t/[{\rm GeV}]-165$.
Here $M_\tau$ and $M_t$  are the physical tau and top quark
masses, where  we take $M_t=174.1$ GeV in our analyses.

The prediction of $\alpha_3(M_Z)$ is shown in Fig.1. 
The four lines correspond to $x=0.00, 0.01, 0.03$ and 0.05.
The uppermost line corresponds to $x=0.00$, while the lowest line 
corresponds to $x=0.05$, that is, the predicted value 
$\alpha_3(M_Z)$ decreases as $x$ increases.
For example, for $x=0.05$ we find a good agreement of $\alpha_3(M_Z)$ 
with the experimental value at $\mu_0=10$ TeV, while we 
obtain a good prediction for $x=0.03$ at $\mu_0=10^{10}$ GeV.
Thus, non-vanishing values of $x$ can lead to the precise value 
$\alpha_3(M_Z)$ even for $\mu_0 \neq 10^{16}$ GeV.
The suitable value of $\mu_0$ is very sensitive to $x$, 
and it changes from a TeV scale to $10^{16}$ GeV when we vary $x$ by 
$O(0.05)$.
Negative values of $x$ lead incorrect $\alpha_3(M_Z)$.
Similarly, we can calculate the case with $\delta >1$.
For larger $\delta$, we find larger $\alpha_3(M_Z)$ as shown 
in Ref.\cite{GR}.

\begin{center}
\setlength{\unitlength}{0.240900pt}
\ifx\plotpoint\undefined\newsavebox{\plotpoint}\fi
\sbox{\plotpoint}{\rule[-0.200pt]{0.400pt}{0.400pt}}%
\begin{picture}(1500,900)(0,0)
\font\gnuplot=cmr10 at 10pt
\gnuplot
\sbox{\plotpoint}{\rule[-0.200pt]{0.400pt}{0.400pt}}%
\put(181.0,123.0){\rule[-0.200pt]{4.818pt}{0.400pt}}
\put(161,123){\makebox(0,0)[r]{0.1}}
\put(1419.0,123.0){\rule[-0.200pt]{4.818pt}{0.400pt}}
\put(181.0,246.0){\rule[-0.200pt]{4.818pt}{0.400pt}}
\put(161,246){\makebox(0,0)[r]{0.11}}
\put(1419.0,246.0){\rule[-0.200pt]{4.818pt}{0.400pt}}
\put(181.0,369.0){\rule[-0.200pt]{4.818pt}{0.400pt}}
\put(161,369){\makebox(0,0)[r]{0.12}}
\put(1419.0,369.0){\rule[-0.200pt]{4.818pt}{0.400pt}}
\put(181.0,492.0){\rule[-0.200pt]{4.818pt}{0.400pt}}
\put(161,492){\makebox(0,0)[r]{0.13}}
\put(1419.0,492.0){\rule[-0.200pt]{4.818pt}{0.400pt}}
\put(181.0,614.0){\rule[-0.200pt]{4.818pt}{0.400pt}}
\put(161,614){\makebox(0,0)[r]{0.14}}
\put(1419.0,614.0){\rule[-0.200pt]{4.818pt}{0.400pt}}
\put(181.0,737.0){\rule[-0.200pt]{4.818pt}{0.400pt}}
\put(161,737){\makebox(0,0)[r]{0.15}}
\put(1419.0,737.0){\rule[-0.200pt]{4.818pt}{0.400pt}}
\put(181.0,860.0){\rule[-0.200pt]{4.818pt}{0.400pt}}
\put(161,860){\makebox(0,0)[r]{0.16}}
\put(1419.0,860.0){\rule[-0.200pt]{4.818pt}{0.400pt}}
\put(181.0,123.0){\rule[-0.200pt]{0.400pt}{4.818pt}}
\put(181,82){\makebox(0,0){4}}
\put(181.0,840.0){\rule[-0.200pt]{0.400pt}{4.818pt}}
\put(391.0,123.0){\rule[-0.200pt]{0.400pt}{4.818pt}}
\put(391,82){\makebox(0,0){6}}
\put(391.0,840.0){\rule[-0.200pt]{0.400pt}{4.818pt}}
\put(600.0,123.0){\rule[-0.200pt]{0.400pt}{4.818pt}}
\put(600,82){\makebox(0,0){8}}
\put(600.0,840.0){\rule[-0.200pt]{0.400pt}{4.818pt}}
\put(810.0,123.0){\rule[-0.200pt]{0.400pt}{4.818pt}}
\put(810,82){\makebox(0,0){10}}
\put(810.0,840.0){\rule[-0.200pt]{0.400pt}{4.818pt}}
\put(1020.0,123.0){\rule[-0.200pt]{0.400pt}{4.818pt}}
\put(1020,82){\makebox(0,0){12}}
\put(1020.0,840.0){\rule[-0.200pt]{0.400pt}{4.818pt}}
\put(1229.0,123.0){\rule[-0.200pt]{0.400pt}{4.818pt}}
\put(1229,82){\makebox(0,0){14}}
\put(1229.0,840.0){\rule[-0.200pt]{0.400pt}{4.818pt}}
\put(1439.0,123.0){\rule[-0.200pt]{0.400pt}{4.818pt}}
\put(1439,82){\makebox(0,0){16}}
\put(1439.0,840.0){\rule[-0.200pt]{0.400pt}{4.818pt}}
\put(181.0,123.0){\rule[-0.200pt]{303.052pt}{0.400pt}}
\put(1439.0,123.0){\rule[-0.200pt]{0.400pt}{177.543pt}}
\put(181.0,860.0){\rule[-0.200pt]{303.052pt}{0.400pt}}
\put(40,441){\makebox(0,0){$\alpha_3(M_Z)$}}
\put(810,21){\makebox(0,0){$log_{10}\mu_0$ [GeV]}}
\put(496,799){\makebox(0,0)[r]{$x=0.00$}}
\put(443,676){\makebox(0,0)[r]{$0.01$}}
\put(443,516){\makebox(0,0)[r]{$0.03$}}
\put(443,246){\makebox(0,0)[r]{$0.05$}}
\put(181.0,123.0){\rule[-0.200pt]{0.400pt}{177.543pt}}
\sbox{\plotpoint}{\rule[-0.400pt]{0.800pt}{0.800pt}}%
\put(181,840){\usebox{\plotpoint}}
\multiput(181.00,838.09)(1.078,-0.502){91}{\rule{1.914pt}{0.121pt}}
\multiput(181.00,838.34)(101.027,-49.000){2}{\rule{0.957pt}{0.800pt}}
\multiput(286.00,789.09)(0.977,-0.502){101}{\rule{1.756pt}{0.121pt}}
\multiput(286.00,789.34)(101.356,-54.000){2}{\rule{0.878pt}{0.800pt}}
\multiput(391.00,735.09)(1.101,-0.502){89}{\rule{1.950pt}{0.121pt}}
\multiput(391.00,735.34)(100.953,-48.000){2}{\rule{0.975pt}{0.800pt}}
\multiput(496.00,687.09)(1.091,-0.502){89}{\rule{1.933pt}{0.121pt}}
\multiput(496.00,687.34)(99.987,-48.000){2}{\rule{0.967pt}{0.800pt}}
\multiput(600.00,639.09)(1.232,-0.502){79}{\rule{2.153pt}{0.121pt}}
\multiput(600.00,639.34)(100.530,-43.000){2}{\rule{1.077pt}{0.800pt}}
\multiput(705.00,596.09)(1.293,-0.502){75}{\rule{2.249pt}{0.121pt}}
\multiput(705.00,596.34)(100.333,-41.000){2}{\rule{1.124pt}{0.800pt}}
\multiput(810.00,555.09)(1.261,-0.502){77}{\rule{2.200pt}{0.121pt}}
\multiput(810.00,555.34)(100.434,-42.000){2}{\rule{1.100pt}{0.800pt}}
\multiput(915.00,513.09)(1.397,-0.503){69}{\rule{2.411pt}{0.121pt}}
\multiput(915.00,513.34)(99.997,-38.000){2}{\rule{1.205pt}{0.800pt}}
\multiput(1020.00,475.09)(1.519,-0.503){63}{\rule{2.600pt}{0.121pt}}
\multiput(1020.00,475.34)(99.604,-35.000){2}{\rule{1.300pt}{0.800pt}}
\multiput(1125.00,440.09)(1.704,-0.503){55}{\rule{2.884pt}{0.121pt}}
\multiput(1125.00,440.34)(98.014,-31.000){2}{\rule{1.442pt}{0.800pt}}
\multiput(1229.00,409.09)(1.911,-0.504){49}{\rule{3.200pt}{0.121pt}}
\multiput(1229.00,409.34)(98.358,-28.000){2}{\rule{1.600pt}{0.800pt}}
\multiput(1334.00,381.09)(3.028,-0.506){29}{\rule{4.867pt}{0.122pt}}
\multiput(1334.00,381.34)(94.899,-18.000){2}{\rule{2.433pt}{0.800pt}}
\put(181,719){\usebox{\plotpoint}}
\multiput(181.00,717.09)(1.232,-0.502){79}{\rule{2.153pt}{0.121pt}}
\multiput(181.00,717.34)(100.530,-43.000){2}{\rule{1.077pt}{0.800pt}}
\multiput(286.00,674.09)(1.397,-0.503){69}{\rule{2.411pt}{0.121pt}}
\multiput(286.00,674.34)(99.997,-38.000){2}{\rule{1.205pt}{0.800pt}}
\multiput(391.00,636.09)(1.261,-0.502){77}{\rule{2.200pt}{0.121pt}}
\multiput(391.00,636.34)(100.434,-42.000){2}{\rule{1.100pt}{0.800pt}}
\multiput(496.00,594.09)(1.422,-0.503){67}{\rule{2.449pt}{0.121pt}}
\multiput(496.00,594.34)(98.918,-37.000){2}{\rule{1.224pt}{0.800pt}}
\multiput(600.00,557.09)(1.436,-0.503){67}{\rule{2.470pt}{0.121pt}}
\multiput(600.00,557.34)(99.873,-37.000){2}{\rule{1.235pt}{0.800pt}}
\multiput(705.00,520.09)(1.519,-0.503){63}{\rule{2.600pt}{0.121pt}}
\multiput(705.00,520.34)(99.604,-35.000){2}{\rule{1.300pt}{0.800pt}}
\multiput(810.00,485.09)(1.614,-0.503){59}{\rule{2.745pt}{0.121pt}}
\multiput(810.00,485.34)(99.302,-33.000){2}{\rule{1.373pt}{0.800pt}}
\multiput(915.00,452.09)(1.614,-0.503){59}{\rule{2.745pt}{0.121pt}}
\multiput(915.00,452.34)(99.302,-33.000){2}{\rule{1.373pt}{0.800pt}}
\multiput(1020.00,419.09)(1.843,-0.504){51}{\rule{3.097pt}{0.121pt}}
\multiput(1020.00,419.34)(98.573,-29.000){2}{\rule{1.548pt}{0.800pt}}
\multiput(1125.00,390.09)(1.965,-0.504){47}{\rule{3.281pt}{0.121pt}}
\multiput(1125.00,390.34)(97.189,-27.000){2}{\rule{1.641pt}{0.800pt}}
\multiput(1229.00,363.09)(2.241,-0.504){41}{\rule{3.700pt}{0.122pt}}
\multiput(1229.00,363.34)(97.320,-24.000){2}{\rule{1.850pt}{0.800pt}}
\multiput(1334.00,339.09)(3.963,-0.509){21}{\rule{6.200pt}{0.123pt}}
\multiput(1334.00,339.34)(92.132,-14.000){2}{\rule{3.100pt}{0.800pt}}
\put(181,517){\usebox{\plotpoint}}
\multiput(181.00,515.09)(2.063,-0.504){45}{\rule{3.431pt}{0.121pt}}
\multiput(181.00,515.34)(97.879,-26.000){2}{\rule{1.715pt}{0.800pt}}
\multiput(286.00,489.09)(1.911,-0.504){49}{\rule{3.200pt}{0.121pt}}
\multiput(286.00,489.34)(98.358,-28.000){2}{\rule{1.600pt}{0.800pt}}
\multiput(391.00,461.09)(2.148,-0.504){43}{\rule{3.560pt}{0.121pt}}
\multiput(391.00,461.34)(97.611,-25.000){2}{\rule{1.780pt}{0.800pt}}
\multiput(496.00,436.09)(2.127,-0.504){43}{\rule{3.528pt}{0.121pt}}
\multiput(496.00,436.34)(96.677,-25.000){2}{\rule{1.764pt}{0.800pt}}
\multiput(600.00,411.09)(2.148,-0.504){43}{\rule{3.560pt}{0.121pt}}
\multiput(600.00,411.34)(97.611,-25.000){2}{\rule{1.780pt}{0.800pt}}
\multiput(705.00,386.09)(2.342,-0.505){39}{\rule{3.852pt}{0.122pt}}
\multiput(705.00,386.34)(97.005,-23.000){2}{\rule{1.926pt}{0.800pt}}
\multiput(810.00,363.09)(2.342,-0.505){39}{\rule{3.852pt}{0.122pt}}
\multiput(810.00,363.34)(97.005,-23.000){2}{\rule{1.926pt}{0.800pt}}
\multiput(915.00,340.09)(2.575,-0.505){35}{\rule{4.200pt}{0.122pt}}
\multiput(915.00,340.34)(96.283,-21.000){2}{\rule{2.100pt}{0.800pt}}
\multiput(1020.00,319.09)(2.453,-0.505){37}{\rule{4.018pt}{0.122pt}}
\multiput(1020.00,319.34)(96.660,-22.000){2}{\rule{2.009pt}{0.800pt}}
\multiput(1125.00,297.09)(2.833,-0.506){31}{\rule{4.579pt}{0.122pt}}
\multiput(1125.00,297.34)(94.496,-19.000){2}{\rule{2.289pt}{0.800pt}}
\multiput(1229.00,278.09)(3.432,-0.507){25}{\rule{5.450pt}{0.122pt}}
\multiput(1229.00,278.34)(93.688,-16.000){2}{\rule{2.725pt}{0.800pt}}
\multiput(1334.00,262.08)(6.543,-0.516){11}{\rule{9.533pt}{0.124pt}}
\multiput(1334.00,262.34)(85.213,-9.000){2}{\rule{4.767pt}{0.800pt}}
\put(181,351){\usebox{\plotpoint}}
\multiput(181.00,349.08)(4.297,-0.509){19}{\rule{6.662pt}{0.123pt}}
\multiput(181.00,349.34)(91.174,-13.000){2}{\rule{3.331pt}{0.800pt}}
\multiput(286.00,336.09)(3.432,-0.507){25}{\rule{5.450pt}{0.122pt}}
\multiput(286.00,336.34)(93.688,-16.000){2}{\rule{2.725pt}{0.800pt}}
\multiput(391.00,320.09)(3.432,-0.507){25}{\rule{5.450pt}{0.122pt}}
\multiput(391.00,320.34)(93.688,-16.000){2}{\rule{2.725pt}{0.800pt}}
\multiput(496.00,304.09)(3.399,-0.507){25}{\rule{5.400pt}{0.122pt}}
\multiput(496.00,304.34)(92.792,-16.000){2}{\rule{2.700pt}{0.800pt}}
\multiput(600.00,288.09)(3.678,-0.508){23}{\rule{5.800pt}{0.122pt}}
\multiput(600.00,288.34)(92.962,-15.000){2}{\rule{2.900pt}{0.800pt}}
\multiput(705.00,273.09)(3.678,-0.508){23}{\rule{5.800pt}{0.122pt}}
\multiput(705.00,273.34)(92.962,-15.000){2}{\rule{2.900pt}{0.800pt}}
\multiput(810.00,258.09)(3.678,-0.508){23}{\rule{5.800pt}{0.122pt}}
\multiput(810.00,258.34)(92.962,-15.000){2}{\rule{2.900pt}{0.800pt}}
\multiput(915.00,243.09)(3.963,-0.509){21}{\rule{6.200pt}{0.123pt}}
\multiput(915.00,243.34)(92.132,-14.000){2}{\rule{3.100pt}{0.800pt}}
\multiput(1020.00,229.09)(3.963,-0.509){21}{\rule{6.200pt}{0.123pt}}
\multiput(1020.00,229.34)(92.132,-14.000){2}{\rule{3.100pt}{0.800pt}}
\multiput(1125.00,215.08)(4.256,-0.509){19}{\rule{6.600pt}{0.123pt}}
\multiput(1125.00,215.34)(90.301,-13.000){2}{\rule{3.300pt}{0.800pt}}
\multiput(1229.00,202.08)(5.775,-0.514){13}{\rule{8.600pt}{0.124pt}}
\multiput(1229.00,202.34)(87.150,-10.000){2}{\rule{4.300pt}{0.800pt}}
\put(1334,190.84){\rule{25.295pt}{0.800pt}}
\multiput(1334.00,192.34)(52.500,-3.000){2}{\rule{12.647pt}{0.800pt}}
\sbox{\plotpoint}{\rule[-0.200pt]{0.400pt}{0.400pt}}%
\put(181,356){\usebox{\plotpoint}}
\put(181.00,356.00){\usebox{\plotpoint}}
\put(201.76,356.00){\usebox{\plotpoint}}
\put(222.51,356.00){\usebox{\plotpoint}}
\put(243.27,356.00){\usebox{\plotpoint}}
\put(264.02,356.00){\usebox{\plotpoint}}
\put(284.78,356.00){\usebox{\plotpoint}}
\put(305.53,356.00){\usebox{\plotpoint}}
\put(326.29,356.00){\usebox{\plotpoint}}
\put(347.04,356.00){\usebox{\plotpoint}}
\put(367.80,356.00){\usebox{\plotpoint}}
\put(388.55,356.00){\usebox{\plotpoint}}
\put(409.31,356.00){\usebox{\plotpoint}}
\put(430.07,356.00){\usebox{\plotpoint}}
\put(450.82,356.00){\usebox{\plotpoint}}
\put(471.58,356.00){\usebox{\plotpoint}}
\put(492.33,356.00){\usebox{\plotpoint}}
\put(513.09,356.00){\usebox{\plotpoint}}
\put(533.84,356.00){\usebox{\plotpoint}}
\put(554.60,356.00){\usebox{\plotpoint}}
\put(575.35,356.00){\usebox{\plotpoint}}
\put(596.11,356.00){\usebox{\plotpoint}}
\put(616.87,356.00){\usebox{\plotpoint}}
\put(637.62,356.00){\usebox{\plotpoint}}
\put(658.38,356.00){\usebox{\plotpoint}}
\put(679.13,356.00){\usebox{\plotpoint}}
\put(699.89,356.00){\usebox{\plotpoint}}
\put(720.64,356.00){\usebox{\plotpoint}}
\put(741.40,356.00){\usebox{\plotpoint}}
\put(762.15,356.00){\usebox{\plotpoint}}
\put(782.91,356.00){\usebox{\plotpoint}}
\put(803.66,356.00){\usebox{\plotpoint}}
\put(824.42,356.00){\usebox{\plotpoint}}
\put(845.18,356.00){\usebox{\plotpoint}}
\put(865.93,356.00){\usebox{\plotpoint}}
\put(886.69,356.00){\usebox{\plotpoint}}
\put(907.44,356.00){\usebox{\plotpoint}}
\put(928.20,356.00){\usebox{\plotpoint}}
\put(948.95,356.00){\usebox{\plotpoint}}
\put(969.71,356.00){\usebox{\plotpoint}}
\put(990.46,356.00){\usebox{\plotpoint}}
\put(1011.22,356.00){\usebox{\plotpoint}}
\put(1031.98,356.00){\usebox{\plotpoint}}
\put(1052.73,356.00){\usebox{\plotpoint}}
\put(1073.49,356.00){\usebox{\plotpoint}}
\put(1094.24,356.00){\usebox{\plotpoint}}
\put(1115.00,356.00){\usebox{\plotpoint}}
\put(1135.75,356.00){\usebox{\plotpoint}}
\put(1156.51,356.00){\usebox{\plotpoint}}
\put(1177.26,356.00){\usebox{\plotpoint}}
\put(1198.02,356.00){\usebox{\plotpoint}}
\put(1218.77,356.00){\usebox{\plotpoint}}
\put(1239.53,356.00){\usebox{\plotpoint}}
\put(1260.29,356.00){\usebox{\plotpoint}}
\put(1281.04,356.00){\usebox{\plotpoint}}
\put(1301.80,356.00){\usebox{\plotpoint}}
\put(1322.55,356.00){\usebox{\plotpoint}}
\put(1343.31,356.00){\usebox{\plotpoint}}
\put(1364.06,356.00){\usebox{\plotpoint}}
\put(1384.82,356.00){\usebox{\plotpoint}}
\put(1405.57,356.00){\usebox{\plotpoint}}
\put(1426.33,356.00){\usebox{\plotpoint}}
\put(1439,356){\usebox{\plotpoint}}
\end{picture}

Fig.1: The prediction of $\alpha_3(M_Z)$
\end{center}

Change of the running behaviour of $g_i$ ($i=1,2,3$) due to $x$ 
affects the running behaviour of other couplings, e.g. 
the running behaviour of the Yukawa couplings.
Furthermore, the Yukawa couplings have corrections due to higher 
dimensional operators.
The bottom and tau Yukawa couplings have the correction term, 
$c'(\Phi/M)({\bf 10})({\bf \bar 5})H_d$, and that changes the 
initial condition of the $b-\tau$ Yukawa unification, 
$g_b(1+C_b)=g_\tau(1+C_\tau)$ with the ratio $C_b/C_\tau=-2/3$.
Naturally, $C_b$ and $C_\tau$ would be of the same order as those 
of the gauge coupling corrections $C_i$, but 
there is no closer relation between  $C_b$ ($C_\tau$) and $C_i$, 
because $c$ and $c'$ couplings are independent of 
each other \footnote{Within the framework of  
gauge-Yukawa unified theories, these couplings could be related 
each other \cite{gYu}.}.

Now let us calculate effects of $x$ and $C_b$ ($C_\tau$) 
on the bottom mass $m_b(M_Z)$ under the $b-\tau$ Yukawa unification 
in the theory with an extra dimension.
The results for $C_b=C_\tau=0$ are shown in Fig. 2.
We have taken $\tan \beta =3$.
The three lines in Fig.2 correspond to $x=0.00$, $0.03$ and 0.05.
The uppermost (lowest) is for $x=0.00$ (0.05).
As $x$ increases, the bottom mass $m_b(M_Z)$ decreases for any value of 
$\mu_0$.
The case with $\mu_0=10$ TeV and $x=0.05$ leads to slightly smaller 
bottom mass than a combination of higher $\mu_0$ and smaller $|x|$ 
leading a good prediction of $\alpha_3(M_Z)$, e.g. 
$(x,\mu_0[{\rm GeV}])=(0.03,10^{10})$ or $(0.00,10^{16})$.

\begin{center}
\setlength{\unitlength}{0.240900pt}
\ifx\plotpoint\undefined\newsavebox{\plotpoint}\fi
\sbox{\plotpoint}{\rule[-0.200pt]{0.400pt}{0.400pt}}%
\begin{picture}(1500,900)(0,0)
\font\gnuplot=cmr10 at 10pt
\gnuplot
\sbox{\plotpoint}{\rule[-0.200pt]{0.400pt}{0.400pt}}%
\put(161.0,123.0){\rule[-0.200pt]{4.818pt}{0.400pt}}
\put(141,123){\makebox(0,0)[r]{2}}
\put(1419.0,123.0){\rule[-0.200pt]{4.818pt}{0.400pt}}
\put(161.0,246.0){\rule[-0.200pt]{4.818pt}{0.400pt}}
\put(141,246){\makebox(0,0)[r]{2.5}}
\put(1419.0,246.0){\rule[-0.200pt]{4.818pt}{0.400pt}}
\put(161.0,369.0){\rule[-0.200pt]{4.818pt}{0.400pt}}
\put(141,369){\makebox(0,0)[r]{3}}
\put(1419.0,369.0){\rule[-0.200pt]{4.818pt}{0.400pt}}
\put(161.0,492.0){\rule[-0.200pt]{4.818pt}{0.400pt}}
\put(141,492){\makebox(0,0)[r]{3.5}}
\put(1419.0,492.0){\rule[-0.200pt]{4.818pt}{0.400pt}}
\put(161.0,614.0){\rule[-0.200pt]{4.818pt}{0.400pt}}
\put(141,614){\makebox(0,0)[r]{4}}
\put(1419.0,614.0){\rule[-0.200pt]{4.818pt}{0.400pt}}
\put(161.0,737.0){\rule[-0.200pt]{4.818pt}{0.400pt}}
\put(141,737){\makebox(0,0)[r]{4.5}}
\put(1419.0,737.0){\rule[-0.200pt]{4.818pt}{0.400pt}}
\put(161.0,860.0){\rule[-0.200pt]{4.818pt}{0.400pt}}
\put(141,860){\makebox(0,0)[r]{5}}
\put(1419.0,860.0){\rule[-0.200pt]{4.818pt}{0.400pt}}
\put(161.0,123.0){\rule[-0.200pt]{0.400pt}{4.818pt}}
\put(161,82){\makebox(0,0){4}}
\put(161.0,840.0){\rule[-0.200pt]{0.400pt}{4.818pt}}
\put(374.0,123.0){\rule[-0.200pt]{0.400pt}{4.818pt}}
\put(374,82){\makebox(0,0){6}}
\put(374.0,840.0){\rule[-0.200pt]{0.400pt}{4.818pt}}
\put(587.0,123.0){\rule[-0.200pt]{0.400pt}{4.818pt}}
\put(587,82){\makebox(0,0){8}}
\put(587.0,840.0){\rule[-0.200pt]{0.400pt}{4.818pt}}
\put(800.0,123.0){\rule[-0.200pt]{0.400pt}{4.818pt}}
\put(800,82){\makebox(0,0){10}}
\put(800.0,840.0){\rule[-0.200pt]{0.400pt}{4.818pt}}
\put(1013.0,123.0){\rule[-0.200pt]{0.400pt}{4.818pt}}
\put(1013,82){\makebox(0,0){12}}
\put(1013.0,840.0){\rule[-0.200pt]{0.400pt}{4.818pt}}
\put(1226.0,123.0){\rule[-0.200pt]{0.400pt}{4.818pt}}
\put(1226,82){\makebox(0,0){14}}
\put(1226.0,840.0){\rule[-0.200pt]{0.400pt}{4.818pt}}
\put(1439.0,123.0){\rule[-0.200pt]{0.400pt}{4.818pt}}
\put(1439,82){\makebox(0,0){16}}
\put(1439.0,840.0){\rule[-0.200pt]{0.400pt}{4.818pt}}
\put(161.0,123.0){\rule[-0.200pt]{307.870pt}{0.400pt}}
\put(1439.0,123.0){\rule[-0.200pt]{0.400pt}{177.543pt}}
\put(161.0,860.0){\rule[-0.200pt]{307.870pt}{0.400pt}}
\put(40,441){\makebox(0,0){$m_b(M_Z)$}}
\put(800,21){\makebox(0,0){$log_{10}\mu_0$ [GeV]}}
\put(481,713){\makebox(0,0)[r]{$x=0.00$}}
\put(481,590){\makebox(0,0)[r]{$0.03$}}
\put(481,467){\makebox(0,0)[r]{$0.05$}}
\put(161.0,123.0){\rule[-0.200pt]{0.400pt}{177.543pt}}
\sbox{\plotpoint}{\rule[-0.400pt]{0.800pt}{0.800pt}}%
\put(161,673){\usebox{\plotpoint}}
\multiput(161.00,674.40)(5.886,0.514){13}{\rule{8.760pt}{0.124pt}}
\multiput(161.00,671.34)(88.818,10.000){2}{\rule{4.380pt}{0.800pt}}
\put(268,680.34){\rule{25.535pt}{0.800pt}}
\multiput(268.00,681.34)(53.000,-2.000){2}{\rule{12.768pt}{0.800pt}}
\put(374,677.84){\rule{25.776pt}{0.800pt}}
\multiput(374.00,679.34)(53.500,-3.000){2}{\rule{12.888pt}{0.800pt}}
\multiput(481.00,676.08)(5.831,-0.514){13}{\rule{8.680pt}{0.124pt}}
\multiput(481.00,676.34)(87.984,-10.000){2}{\rule{4.340pt}{0.800pt}}
\multiput(587.00,666.08)(4.785,-0.511){17}{\rule{7.333pt}{0.123pt}}
\multiput(587.00,666.34)(91.779,-12.000){2}{\rule{3.667pt}{0.800pt}}
\multiput(694.00,654.09)(3.714,-0.508){23}{\rule{5.853pt}{0.122pt}}
\multiput(694.00,654.34)(93.851,-15.000){2}{\rule{2.927pt}{0.800pt}}
\multiput(800.00,639.09)(3.280,-0.507){27}{\rule{5.235pt}{0.122pt}}
\multiput(800.00,639.34)(96.134,-17.000){2}{\rule{2.618pt}{0.800pt}}
\multiput(907.00,622.09)(3.249,-0.507){27}{\rule{5.188pt}{0.122pt}}
\multiput(907.00,622.34)(95.232,-17.000){2}{\rule{2.594pt}{0.800pt}}
\multiput(1013.00,605.09)(2.501,-0.505){37}{\rule{4.091pt}{0.122pt}}
\multiput(1013.00,605.34)(98.509,-22.000){2}{\rule{2.045pt}{0.800pt}}
\multiput(1120.00,583.09)(2.737,-0.505){33}{\rule{4.440pt}{0.122pt}}
\multiput(1120.00,583.34)(96.785,-20.000){2}{\rule{2.220pt}{0.800pt}}
\multiput(1226.00,563.09)(2.916,-0.506){31}{\rule{4.705pt}{0.122pt}}
\multiput(1226.00,563.34)(97.234,-19.000){2}{\rule{2.353pt}{0.800pt}}
\multiput(1333.00,544.08)(4.338,-0.509){19}{\rule{6.723pt}{0.123pt}}
\multiput(1333.00,544.34)(92.046,-13.000){2}{\rule{3.362pt}{0.800pt}}
\put(161,558){\usebox{\plotpoint}}
\multiput(161.00,559.40)(5.886,0.514){13}{\rule{8.760pt}{0.124pt}}
\multiput(161.00,556.34)(88.818,10.000){2}{\rule{4.380pt}{0.800pt}}
\put(268,567.34){\rule{25.535pt}{0.800pt}}
\multiput(268.00,566.34)(53.000,2.000){2}{\rule{12.768pt}{0.800pt}}
\put(481,567.34){\rule{25.535pt}{0.800pt}}
\multiput(481.00,568.34)(53.000,-2.000){2}{\rule{12.768pt}{0.800pt}}
\multiput(587.00,566.08)(7.719,-0.520){9}{\rule{10.900pt}{0.125pt}}
\multiput(587.00,566.34)(84.377,-8.000){2}{\rule{5.450pt}{0.800pt}}
\multiput(694.00,558.08)(9.147,-0.526){7}{\rule{12.314pt}{0.127pt}}
\multiput(694.00,558.34)(80.441,-7.000){2}{\rule{6.157pt}{0.800pt}}
\multiput(800.00,551.08)(5.886,-0.514){13}{\rule{8.760pt}{0.124pt}}
\multiput(800.00,551.34)(88.818,-10.000){2}{\rule{4.380pt}{0.800pt}}
\multiput(907.00,541.08)(4.740,-0.511){17}{\rule{7.267pt}{0.123pt}}
\multiput(907.00,541.34)(90.918,-12.000){2}{\rule{3.633pt}{0.800pt}}
\multiput(1013.00,529.09)(3.749,-0.508){23}{\rule{5.907pt}{0.122pt}}
\multiput(1013.00,529.34)(94.740,-15.000){2}{\rule{2.953pt}{0.800pt}}
\multiput(1120.00,514.08)(4.740,-0.511){17}{\rule{7.267pt}{0.123pt}}
\multiput(1120.00,514.34)(90.918,-12.000){2}{\rule{3.633pt}{0.800pt}}
\multiput(1226.00,502.09)(3.749,-0.508){23}{\rule{5.907pt}{0.122pt}}
\multiput(1226.00,502.34)(94.740,-15.000){2}{\rule{2.953pt}{0.800pt}}
\multiput(1333.00,487.08)(9.147,-0.526){7}{\rule{12.314pt}{0.127pt}}
\multiput(1333.00,487.34)(80.441,-7.000){2}{\rule{6.157pt}{0.800pt}}
\put(374.0,570.0){\rule[-0.400pt]{25.776pt}{0.800pt}}
\put(161,494){\usebox{\plotpoint}}
\multiput(161.00,495.40)(5.886,0.514){13}{\rule{8.760pt}{0.124pt}}
\multiput(161.00,492.34)(88.818,10.000){2}{\rule{4.380pt}{0.800pt}}
\multiput(268.00,505.38)(17.384,0.560){3}{\rule{17.160pt}{0.135pt}}
\multiput(268.00,502.34)(70.384,5.000){2}{\rule{8.580pt}{0.800pt}}
\put(374,508.34){\rule{25.776pt}{0.800pt}}
\multiput(374.00,507.34)(53.500,2.000){2}{\rule{12.888pt}{0.800pt}}
\put(481,508.34){\rule{25.535pt}{0.800pt}}
\multiput(481.00,509.34)(53.000,-2.000){2}{\rule{12.768pt}{0.800pt}}
\put(587,505.84){\rule{25.776pt}{0.800pt}}
\multiput(587.00,507.34)(53.500,-3.000){2}{\rule{12.888pt}{0.800pt}}
\multiput(694.00,504.06)(17.384,-0.560){3}{\rule{17.160pt}{0.135pt}}
\multiput(694.00,504.34)(70.384,-5.000){2}{\rule{8.580pt}{0.800pt}}
\multiput(800.00,499.08)(9.235,-0.526){7}{\rule{12.429pt}{0.127pt}}
\multiput(800.00,499.34)(81.204,-7.000){2}{\rule{6.214pt}{0.800pt}}
\multiput(907.00,492.08)(5.831,-0.514){13}{\rule{8.680pt}{0.124pt}}
\multiput(907.00,492.34)(87.984,-10.000){2}{\rule{4.340pt}{0.800pt}}
\multiput(1013.00,482.08)(5.886,-0.514){13}{\rule{8.760pt}{0.124pt}}
\multiput(1013.00,482.34)(88.818,-10.000){2}{\rule{4.380pt}{0.800pt}}
\multiput(1120.00,472.08)(5.831,-0.514){13}{\rule{8.680pt}{0.124pt}}
\multiput(1120.00,472.34)(87.984,-10.000){2}{\rule{4.340pt}{0.800pt}}
\multiput(1226.00,462.08)(6.669,-0.516){11}{\rule{9.711pt}{0.124pt}}
\multiput(1226.00,462.34)(86.844,-9.000){2}{\rule{4.856pt}{0.800pt}}
\multiput(1333.00,453.06)(17.384,-0.560){3}{\rule{17.160pt}{0.135pt}}
\multiput(1333.00,453.34)(70.384,-5.000){2}{\rule{8.580pt}{0.800pt}}
\end{picture}

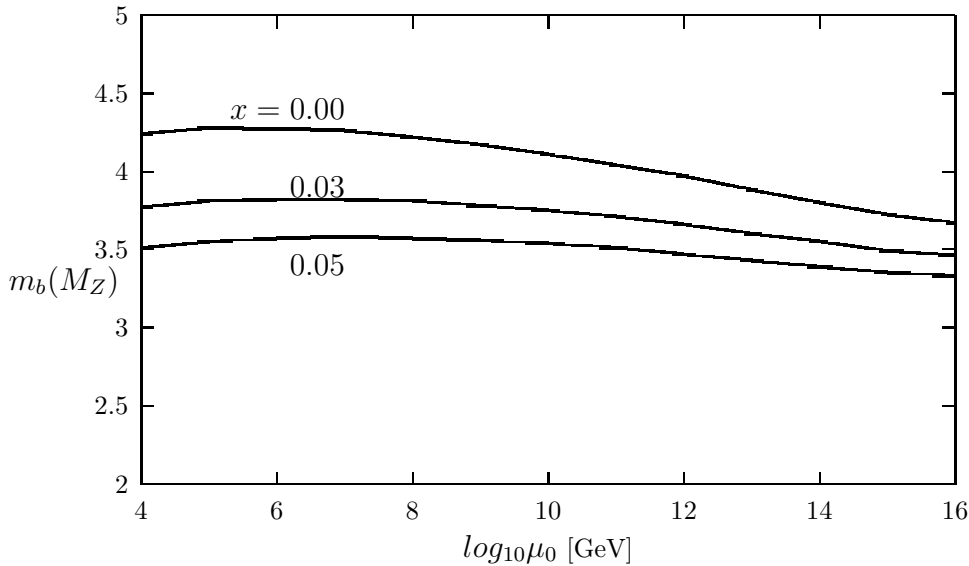
Fig.2:The bottom mass 
\end{center}

Similarily, we can discuss the case with non-vanishing $C_b$ and $C_\tau$.
As an example, let us consider the initial condition that 
$g_b$ is less than $g_\tau$ by $3\%$.
In this case, we have $m_b(M_Z)=3.40$ GeV for $\mu_0=10$ TeV and 
$x=0.05$ and $m_b(M_Z)=3.65$ GeV for $\mu_0=10^{10}$ GeV and 
$x=0.03$.
These values are different from the corresponding values in Fig.2 
by $3\%$.
Thus, if we introduce a few percentage of the difference for 
the $b-\tau$ Yukawa unification, the predicted values of $m_b(M_Z)$ 
are shifted from values in Fig.2 by almost same percentage.

Similarily, we can discuss the case with large $\tan \beta$.
Indeed, the case with $\tan \beta=20$ has a 
behaviour very similar to Fig. 2.
Furthermore, much larger $\tan \beta$ leads to smaller $m_b(M_Z)$ as 
shown in Ref.~\cite{KKMZ}.
However, in the large $\tan \beta$ we have sizable SUSY corrections 
to the bottom mass \cite{hall,COPW}. 
They depend on superparticle masses and increase as $\tan \beta$ 
incerases.
For example, the case with $\tan \beta=20$ has $O(10 \%)$ of 
SUSY corrections to the bottom mass.
Thus, the detailed prediction of the bottom mass for large 
$\tan \beta$ is beyond our scope.

To summarize, we have studied the effects due to 
the higher dimensional operators on the gauge coupling unification 
as well as the running behaviour of the Yukawa couplings.
The energy scale $\mu_0$ leading to a good prediction of $\alpha_3(M_Z)$ 
is very sensitive to $x$ even for $x =O(0.01)$.
For example, in the case with  $x=0.05$ the gauge coupling unification can be 
realized for $\mu_0=O(10)$ TeV.

{\bf Note added}

After completion of this work, an article \cite{CDH} appeared, 
where effects due to higher dimensional operators to the 
gauge couplings are also discussed.

{\bf Acknowledgements}

The authors would like to thank J.~Kubo, M.~Mondrag\' on
and G.~Zoupanos for their encouragements and useful discussions.


\begin{thebibliography}{99}

\bi{antoniadis4}
See for early works, \\
I.~Antoniadis, Phys. Lett. {\bf B246} (1990) 377; 
I.~Antoniadis, C.~Mu\~noz and M. Quir\'os,  
Nucl. Phys. {\bf B397} (1983) 515; 
I.~Antoniadis and K.~Benakli, Phys. Lett. {\bf B326} (1994) 69; 
I.~Antoniadis, K.~Benakli and M. Quir\'os, 
Phys. Lett. {\bf B331} (1994) 313.


\bi{witten1}
E. Witten, Nucl. Phys. {\bf B471} (1996) 135;
P. Horava and E. Witten, Nucl. Phys. {\bf B475} (1996) 94; 
J.~Lykken, Phys. Rev. {\bf D} (1996) 3693.


\bi{arkani1}
N. Arkani-Hamed, S. Dimopoulos and G. Dvali,
Phys. Lett. {\bf B429} (1998) 263; 
Phys. ReV. {\bf D59} (1999) 086004;
I. Antoniadis, N. Arkani-Hamed, S. Dimopoulos and G. Dvali, 
Phys. Lett. {\bf B436} (1998) 257.

\bi{extra-ref}
See also, e.g. 
N. Arkani-Hamed, S. Dimopoulos and J. March-Russell,
hep-th/9809124; 
I. Antoniadis, S. Dimopoulos, A. Pomarol and M. Quir\'os,
Nucl. Phys. {\bf B544} (1999) 503; 
N. Arkani-Hamed and S. Dimopoulos,
hep-ph/9811353; 
Z. Berezhiani and G. Dvali,
Phys. Lett. {\bf B450} (1999) 24; 
N. Arkani-Hamed, S. Dimopoulos, G. Dvali and J. March-Russell,
hep-ph/9811448; 
R. Sundrum,  Phys. Rev. {\bf D 59} (1999) 085009; 
G. Shiu and S.-H.H. Tye,
Phys. Rev. {\bf 58} (1998) 106007; 
A. Pomarol and M. Quir\'os,
Phys. Lett. {\bf B438} (1998) 255; 
C. Bachas,
JHEP {\bf 9811} (1999) 023; 
Z. Kakushadze and S.-H. Tye,
Phys. Rev. {\bf D58} (1998) 126001; 
Nucl. Phys. {\bf B548} (1999) 180;
K. Benakli,  
hep-ph/9809582;
K. Dienes, E. Dudas, T. Gherghetta and A. Riotto, 
Nucl. Phys. {\bf B543} (1999) 387;
L. Randall and R. Sundrum, 
hep-th/9810155;
C.P. Burgess, L.E. Ib\'a\~nez and F. Quevedo,
hep-ph/9810535; 
K.R. Dienes, E. Dudas and T. Gherghetta,
hep-ph/9811428;
Z. Kakushadze, 
hep-th/9811193; hep-th/9812163; hep-th/9902080
H. Hatanaka, T. Inami and C.S. Lim,
Mod. Phys. Lett. {\bf A13} (1998) 2601;
I.¨Antoniadis and C.~Bachas, hep-th/9812093;
L.E. Ib\'a\~nez, C.~Mu\~noz and S.~Rigolin, hep-th/9812397;
A.~Delgado, A.~Pomarol and M.~Quir\'os,
hep-ph/9812489;
T.E. Clark  and S.T. Love,
hep-th/9901103;
A. Donini, S. Rigolin; Nucl. Phys. {\bf B550} (1999) 59;
T. Banks, M. Dine and A. Nelson; hep-th/9903019; 
G.~Shiu, R.~Shrock and S.-H.H.~Tye, 
hep-ph/9904262.




\bi{extra-ref2}
See e.g. 
G.F. Giudice, R. Rattazzi and J.D. Wells,
Nucl. Phys. {\bf B544} (1999) 3;
S. Nussinov and R. Shrock,
Phys. Rev. {\bf D59} (1999) 105002;
E.A. Mirabelli, M. Perelstein and M.E. Peskin,
Phys. Rev. Lett. {82} (1999) 2236;
T. Han, J.D. Lykken and R.J. Zhang,
Phys. Rev. {\bf D59} (1999) 105006;
J. L. Hewett, 
hep-ph/9811356;
P.~Mathews, S.~Raychaudhuri and K.~Sridhar, 
Phys. Lett. {\bf B343} (1999) 343; hep-ph/9812486; hep-ph/9904232;
B.A.~Dobrescu, hep/9812349; hep-ph/9903407;
T.~Rizzo, Phys. Rev. {\bf 59} (1999) 115010; 
hep-ph/9903475; hep-ph/9904380; 
A. E. Faraggi and M. Pospelov,hep-ph/9901299;
K. Agashe and N.G. Deshpande; hep-ph/9902263;
M.L. Graesser, hep-ph/9902310; 
P.~Nath and M.~Yamaguchi, hep-ph/9902323; hep-ph/9903298;
M.~Masip and A.~Pomarol, hep-ph/9902467;
K.~Cheung and W.Y.~Keung, hep-ph/9903294;
S.~Cullen and M.~Perelstein, hep-ph/9903422;
D. Atwood, S. Bar-Shalom and A. Soni,hep-ph/9903538; hep-ph/9906400;
C. Bal\'azs, H.-J. He, W.W. Repko, C.P. Yuan and D.A. Dicu; 
hep-ph/9904220;
K.~Cheung, hep-ph/9904266;
X.~He, hep-ph/9905295;
I.~Antoniadis, K.~Benakli and M. Quir\'os, hep-ph/9905311;
P.~Mathews, P.~Poulose and K.~Sridhar, 
hep-ph/9905395;
T.G.~Rizzo and J.D.~Wells, hep-ph/9906234.





\bi{dienes1}
K. Dienes, E. Dudas and T. Gherghetta,
Phys. Lett. {\bf B436} (1998) 55; 
Nucl. Phys. {\bf B537} (1999) 47.

\bi{GR}
D.~Ghilencea and G.G.~Ross, Phys. Lett. {\bf B442} (1998) 165.


\bi{KKMZ}
T.¨Kobayashi, J.~Kubo, M.~Mondragon and G.¨Zoupanos, 
Nucl. Phys. {\bf B550} (1999) 99.

\bi{p-data}
Reviews of Particle Physics, Particle Data Group, 
Eur. Phys. J. {\bf C3} (1998) 1.


\bi{extra-uni}
C.~Carone, hep-ph/9902407;
A.~Delgado and M.~Quir\'os, hep-ph/9903400; 
P.H.¨Frampton and A.~Rasin, hep-ph/9903479; 
A.~Perez-Lorenzana and R.N.~Mohapatra, hep-ph/9904504.


\bi{AK}
S.A.~Abel and S.F.~King, Phys. Rev. {\bf D59} (1999) 095010.

\bibitem{g-nonuni}
C.T.~Hill, Phys.~Lett., {\bf 135B} (1984) 47;\\
Q.~Shafi and C.~Wetterich, Phys.~Rev.~Lett. {\bf 52} (1984) 875;\\
L.J.~Hall and U.~Sarid,  Phys.~Rev.~Lett. {\bf 70} (1993) 2673.


\bibitem{HKKP}
K.~Huitu, Y.~Kawamura, T.~Kobayashi and K.~Puolam\"aki, in preparation.

\bibitem{jack3} I. Jack and D.R.T. Jones,
Phys. Lett. {\bf B415} (1997) 383; 
I. Jack, D.R.T. Jones and
 A. Pickering, Phys. Lett. {\bf B426} (1998) 73;\\
L.V. Avdeev, 
D.I. Kazakov and I.N. Kondrashuk,
Nucl. Phys. {\bf B510} (1998) 289. 

\bi{kkz1}
See also \\
T. Kobayashi, J. Kubo and G. Zoupanos,
Phys. Lett. {\bf B427} (1998) 291; \\
I. Jack, D.R.T. Jones and
 A. Pickering, Phys. Lett. {\bf B432} (1998) 114;\\
N. Arkani-Hamed, G.F. Giudice, M.A. Luty and R. Rattazzi, 
hep-ph/9803290.

\bi{gYu}
D. Kapetanakis, M. Mondrag\'on and G. Zoupanos,
Zeit.~f.~Phys.~{\bf C60} (1993) 181; 
M. Mondrag\'on and G. Zoupanos, 
Nucl.~Phys.~B (Proc.~Suppl.) {\bf 37C} (1995) 98;
J. Kubo, M. Mondrag\'on and G. Zoupanos, 
Nucl.~Phys.~{\bf B424} (1994) 291; 
T. Kobayashi, J. Kubo, M. Mondrag\'on and G. Zoupanos, 
Nucl.~Phys.~{\bf B511} (1998) 45 and references therein.


\bibitem{hall}
L.J.~Hall, R.~Rattazzi and U.~Sarid, 
Phys. Rev. {\bf D50} (1994) 7048; 
R.~Hempfling, Phys. Rev. {\bf D49} (1994) 6168.

\bibitem{COPW}
M.~Carena, M.~Olechowski, S.~Pokorski and C.E.M.~Wagner, 
Nucl. Phys. {\bf B426} (1994) 269.




\bi{CDH}
H.C.~Cheng, B.A.~Dobrescu and C.T.~Hill, hep-ph/9906327.


\end{thebibliography}
\end{document}